\documentclass[a4paper]{article}
\usepackage{amsmath,amsthm,graphicx,paralist}

\title{A Proof of the Pumping Lemma for Context-Free Languages Through Pushdown
Automata}
\author{Antoine Amarilli\footnote{\'Ecole normale sup\'erieure, Paris, France.}
\and Marc Jeanmougin\footnotemark[\value{footnote}]}

\DeclareMathOperator*{\fp}{fp}
\DeclareMathOperator*{\lp}{lp}
\newcommand{\fpp}[1]{\widehat{\fp{#1}}}
\newcommand{\lpp}[1]{\widehat{\lp{#1}}}

\newtheorem{theorem}{Theorem}

\makeatletter
\newcommand*{\defeq}{\mathrel{\rlap{%
  \raisebox{0.3ex}{$\m@th\cdot$}}%
  \raisebox{-0.3ex}{$\m@th\cdot$}}%
  =}
\makeatother
\begin{document}

\maketitle

\begin{abstract}
The pumping lemma for context-free languages is a result about pushdown
automata which is strikingly similar to the well-known pumping lemma for
regular languages. However, though the lemma for regular languages is simply
proved by using the pigeonhole principle on deterministic automata, the lemma
for pushdown automata is proven through an equivalence with context-free
languages and through the more powerful Ogden's lemma. We present here a proof
of the pumping lemma for context-free languages which relies on pushdown
automata instead of context-free grammars.
\end{abstract}

\section{Setting}

The pumping lemma for regular languages is the following well-known result:
\begin{theorem}
Let $L$ be a regular language over an alphabet $\Sigma$. There exists some
integer $p \geq 1$ such that, for every $w \in L$ such that $|w| > p$, there
exists a decomposition $w = x y z$ such that:
\begin{enumerate}
\item $|xy| \leq p$
\item $|y| \geq 1$
\item $\forall n \geq 0, x y^n z \in L$
\end{enumerate}
\end{theorem}

The pumping lemma for context-free
languages~\cite{BarHillelPerlesShamir61Formal}, also known as the Bar-Hillel
lemma, is the following similar result:

\begin{theorem}
Let $L$ be a context-free language over an alphabet $\Sigma$. There exists some
integer $p \geq 1$ such that, for every $w \in L$ such that $|w| > p$, there
exists a decomposition $w = u v x y z$ such that:
\begin{enumerate}
\item $|vxy| \leq p$
\item $|vy| \geq 1$
\item $\forall n \geq 0, u v^n x y^n z \in L$
\end{enumerate}
\end{theorem}

One would expect the classical proofs of these results to be similar. However,
this is not the case. The pumping lemma for regular languages~\cite{Hopcroft79}
is usually proved through the equivalence between regular languages and finite
automata by picking a deterministic automaton $A$ which recognizes the language
$L$ ; we can then use the fact that the accepting path of any word $w$ longer
than the number of states of $A$ must pass by the same state twice (by the
pigeonhole principle), yielding the points at which we can decompose $w$. The
pumping lemma for context-free languages, however, is usually derived from
Ogden's lemma~\cite{Ogden68} which is itself proved by examining context-free
grammars (CFGs) and not pushdown automata (using the equivalence of these two
formalisms).

It seems reasonable to hope that the pumping lemma for context-free languages
can be proved directly from the properties of pushdown automata, with no
reference to CFGs. In the next section, we propose such a
proof. Though the underlying ideas that we introduce in this proof are
apparently part of the folklore, we are not aware of any attempt to prove the
pumping lemma directly through pushdown automata. The most relevant
existing work that we know of is a weaker form of the
result~\cite{Kartzow12}.

Analogous techniques to the one used below can be used to obtain a proof of
Ogden's lemma. However, it seems that the most natural way to do so is very
similar to a combination of the usual pushdown system encoding to CFGs and the
usual proof of Ogden's lemma. These further efforts (not included in this note)
suggest that the proof below, though it does not mention CFGs on the surface,
may not differ very much from a CFG-based argument after all.

\section{Proof}

Let $L$ be a context-free language over an alphabet $\Sigma$. Let $A$ be a
pushdown automaton which recognizes $L$, with stack alphabet $\Gamma$. We denote
by $|A|$ the number of states of $A$. To simplify the reasoning, we will impose
the following condition on $A$ (denoted by (*)): all transitions of $A$ pop the
topmost symbol of the stack and either push no symbol on the stack or push on
the stack the previous topmost symbol and some other symbol. It is easy to see
that any pushdown automata which pushes arbitrary sequences of symbols on the
stack can be rewritten in this fashion by replacing its transitions by an
initial pop transition followed by a sequence of $\epsilon$-transitions pushing
the appropriate symbols on the stack. (However, keep in mind that because of
this translation, $|A|$ in what follows does not refer to the number of states
of the original automaton recognizing $A$ but to that of its translation by this
process.)

We define $p' = |A|^2 |\Gamma|$ and define the pumping length to be $p = |A|
(|\Gamma|+1)^{p'}$. We will now show that all $w \in L$ such that $|w| > p$ have
a decomposition of the form $w = u v x y z$ such that $|vxy| \leq p$, $|vy| \geq
1$ and $\forall n \geq 0, u v^n x y^n z \in L$.

Let $w \in L$ such that $|w| > p$. Let $\pi$ be an accepting path of minimal
length for $w$ (represented as a sequence of transitions of $A$), we denote its
length by $|\pi|$. We can define, for $0 \leq i < |\pi|$, $s_i$ the size of the
stack at position $i$ of the accepting path. For all $N > 0$, we will define an
\textbf{$N$-level} over $\pi$ as a set of three indices $i, j, k$ with $0 \leq i
< j < k \leq p$ such that the stack grows by $N$ symbols between $i$ and $j$ and
shrinks by $N$ symbols between $j$ and $k$. Formally, we require that:

\begin{enumerate}
  \item $s_i = s_k, s_j = s_i + N$
  \item for all $n$ such that $i \leq n \leq j$, $s_i \leq s_n \leq s_j$ 
  \item for all $n$ such that $j \leq n \leq k$, $s_k \leq s_n \leq s_k$.
\end{enumerate}

We define the level $l$ of $\pi$ as the maximal $N$ such that $\pi$ has an
$N$-level. This definition is motivated by the following observation: if the
size of the stack over a path $\pi$ becomes larger than its level $l$, then the
stack symbols more than $l$ levels deep will never be popped. Formally, we
define the \textbf{configurations} of $A$ as the couples of a state of $A$ and a
sequence of $l$ stack symbols (where stacks of size less than $l$ are
represented by padding them to $l$ with a special blank symbol, which is why we
use $|\Gamma| + 1$ when defining $p$). By definition, there are $|A| (|\Gamma| +
1)^l$ such configurations. Essentially, $A$ acts as a finite automaton without
stack between the configurations.

We can now distinguish two cases: either the level is low and the number of
configurations is small, or the level is high. Formally:

\begin{enumerate}
  \item $l < p'$ and, by the pigeonhole principle, the same configuration is
    encountered twice in the first $p+1$ steps of $\pi$,
  \item $l \geq p'$ and, by the pigeonhole principle, we will prove that a
    certain notion of \emph{full state} is repeated for two different stack
    sizes in any $l$-level of $w$.
\end{enumerate}

\paragraph{Case 1.} $l < p'$. In this case, the number of configurations is less
than $p$. Hence, in the $p+1$ first steps of $\pi$, the same configuration is
encountered twice at two different positions, say $i < j$. Denote by
$\widehat{i}$ (resp. $\widehat{j}$) the position of the last letter of $w$ read
at step $i$ (resp. $j$) of $\pi$. We have $\widehat{i} \leq \widehat{j}$. Hence,
we can factor $w = u v x y z$ with $y z = \epsilon$, $u = w_{0 \cdots
\widehat{i}}$, $v = w_{\widehat{i} \cdots \widehat{j}}$, $x = w_{\widehat{j}
\cdots |w|}$. (By $w_{x \cdots y}$ we denote the letters of $w$ from $x$
inclusive to $y$ exclusive.) By construction, $|vxy| \leq p$.

We also have to show that $\forall n \geq 0, u v^n x y^n z = u v^n x \in L$, but
this follows from our observation above: stack symbols deeper than $l$ are never
popped, so there is no way to distinguish configurations which are equal
according to our definition, and an accepting path for $u v^n x$ is built from
that of $w$ by repeating the steps between $i$ and $j$, $n$ times.

Finally, we also have $|v| > 0$, because if $v = \epsilon$, then, because we
have the same configuration at steps $i$ and $j$ in $\pi$, $\pi' = \pi_{0 \cdots
i} \pi_{j \cdots |\pi|}$ would be an accepting path for $w$, contradicting the
minimality of $\pi$.

\begin{figure}
  \centering
  \label{fig}
  \includegraphics[scale=2]{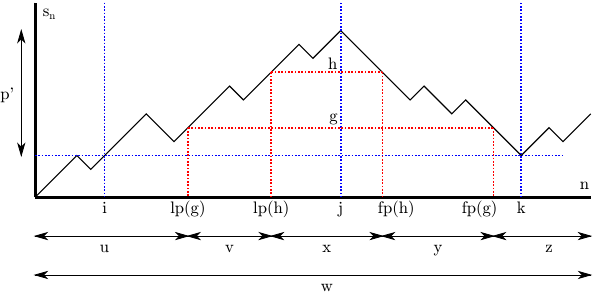}
  \caption{Illustration of the construction for case 2. To simplify the drawing,
  the distinction between the path positions and word positions are omitted.}
\end{figure}

\paragraph{Case 2.} $l \geq p'$. Let $i, j, k$ be a $p'$-level. To any stack
size $h$, $s_i \leq h \leq s_j$, we associate the \textbf{last push}
$\lp(h) = \max(\{y \leq j | s_y = h\})$ and the \textbf{first pop}
$\fp(h) = \min(\{y \geq j | s_y = h\})$.
By definition, $i \leq \lp(h) \leq j$ and $j \leq \fp(h) \leq
k$. We say that the \textbf{full state} of a stack size $h$ is the triple formed
by:

\begin{enumerate}
  \item the automaton state at position $\lp(h)$
  \item the topmost stack symbol at position $\lp(h)$ (which, by construction,
    is also the topmost stack symbol at position $\fp(h)$
  \item the automaton state at position $\fp(h)$
\end{enumerate}

(Observe that there is a link between this definition and what is known as
``Ginsburg triples'' when encoding pushdown systems in CFGs.)

There are $p'$ possible full states, and $p' + 1$ stack sizes between $s_i$ and
$s_j$, so, by the pigeonhole principle, there exist two stack sizes $g, h$ with
$s_i \leq g < h \leq s_j$ such that the full states at $g$ and $h$ are the same.
Like in Case 1, we define by $\lpp(g)$, $\lpp(h)$, $\fpp(h)$ and $\fpp(g)$ the
positions of the last letters of $w$ read at the corresponding positions in $\pi$.
We factor $w = u v x y z$ where $u = w_{0 \cdots \lpp(g)}$,
$v = w_{\lpp(g) \cdots \lpp(h)}$,
$x = w_{\lpp(h) \cdots \fpp(h)}$,
$y = w_{\fpp(h) \cdots \fpp(g)}$,
and $z = w_{\fpp(g) \cdots |w|}$.

This factorization ensures that $|vxy| \leq p$ (because $k \leq p$ by our
definition of levels).

We also have to show that $\forall n \geq 0, u v^n x y^n z \in L$. To do so,
observe that each time that we repeat $v$, we start from the same state and the
same stack top and we do not pop below our current position in the stack
(otherwise we would have to push again at the current position, violating the
maximality of $\lp(g)$), so we can follow the same path in $A$ and push the
same symbol sequence on the stack. By the maximality of $\lp(h)$ and the
minimality of $\fp(h)$, while reading $x$, we do not pop below our current
position in the stack, so the path followed in the automaton is the same
regardless of the number of times we repeated $v$. Now, if we repeat $w$ as many
times as we repeat $v$, since we start from the same state, since we have pushed
the same symbol sequence on the stack with our repeats of $v$, and since we do
not pop more than what $v$ has stacked by minimality of $\fp(g)$, we can follow
the same path in $A$ and pop the same symbol sequence from the stack. Hence, an
accepting path from $u v^n x y^n z$ can be constructed from the accepting path
for $w$.

Finally, we also have $|vy| > 1$, because like in case 1, if $v =
\epsilon$ and $y = \epsilon$, we can build a shorter accepting path for $w$ by
removing $\pi_{\lp(g)\cdots\lp(h)}$ and $\pi_{\fp(h)\cdots\fp(g)}$.\\

Hence, we have an adequate factorization in both cases, and the result is
proved.

\section*{Acknowledgements}

We are grateful to Alexander Kartzow for reporting an error in a previous
version of this proof and for pointing us to relevant work.

\bibliographystyle{alpha}
\bibliography{proof}

\end{document}